\documentclass[vecphys]{svmult}

\usepackage{makeidx}         
\usepackage{graphicx}        
\usepackage{multicol}        

\makeindex             

\begin{document}

\title*{Computing solar absolute fluxes}
\author{Carlos Allende Prieto}
\institute{McDonald Observatory and Department of Astronomy, 
The University of Texas, Austin, Texas 78712, USA
\texttt{callende@astro.as.utexas.edu}
}
%
%
\maketitle

\begin{abstract}
Computed color indices and spectral shapes for individual stars 
are routinely compared with observations for essentially all spectral 
types, but absolute fluxes are rarely tested. 
We can confront observed irradiances with the predictions from model 
atmospheres for a few stars with  accurate angular diameter 
measurements, notably the Sun. Previous calculations have been 
hampered by inconsistencies
and the use of outdated atomic data and abundances. I provide here a 
progress report on our current efforts to compute absolute fluxes 
for  solar model photospheres.
Uncertainties in the solar composition constitute a significant source
of error in computing solar radiative fluxes. 
\end{abstract}

\section{Introduction}
\label{sec:1}

The spectrum of the Sun is the outcome of the physics
governing the outer layers of our star.
Understanding the formation of the solar spectrum is a necessary 
step in order to be able to predict its variability along the solar
magnetic cycle and to measure the solar surface composition. 
The solar spectrum at
wavelengths longer than about 140 nm is variable 
only at the few-percent level, and given
the exquisite accuracy of the solar parameters, 
observations of the Sun may provide the best
available standard to calibrate and guide the construction of theoretical 
model atmospheres for late-type stars. 
Ultimately, our ability to predict the luminosities
of other stars and entire galaxies can be tested and improved by
studying the solar spectrum.

The UV part of the spectrum is of particular relevance for us, 
as it is closely connected to the chemistry of the Earth's
atmosphere, and  the evolution of life on Earth. 
Astrophysically, the UV  is exciting for its wealth of 
information: the strongest atomic lines concentrate
in this spectral window, and so do ionization edges. 
Although the Sun is not a particularly luminous star, it shares 
atmospheric physics with other F-G-K late-type stars which 
contribute significant mass and light to distant galaxies, 
as shown in many of the papers included in this volume.

Perhaps the most severe difficulty to model the outer layers of the
Sun is related to the existence of an 'upper atmosphere', 
where the time-averaged thermal gradient is reversed and a combination 
of high temperature and low density drives the plasma far from 
equilibrium conditions (see, Judge 2005 and Rutten 2007 for recent reviews).
Semi-empirical time-independent models of the upper atmosphere have provided 
significant insight (see, e.g., the classical 
Vernazza, Avrett \& Loesser 1981 paper).  Increasingly 
sophisticated hydrodynamical simulations 
are making their way upwards into the lower chromosphere 
(Wedemeyer et al. 2004; Wedemeyer-B\"ohm et al. 2007).

Space imagery of the upper atmosphere reveals a complicated 
interaction of magnetic field and waves. Such images contrast 
with the much simpler picture 
that we get from optical observations of the photosphere, where
the magnetic field that permeates our
star causes only a small distortion from  {\it field-free} conditions,
and the temperature contrast of the granulation is only a few percent. 
Fortunately, it is possible to study the lower atmosphere independently 
from higher layers. In the optical and infrared the upper atmosphere is
optically thin and the opacity,
dominated by the H$^{-}$ continuum, is only superseded by
metal opacity at wavelengths shorter than about 300 nm. As we move
further into the UV, the rapidly increasing metal opacity 
 shifts the spectrum formation into the lower chromosphere. 
 The change of character is
reflected in the time variability of the integrated solar spectrum, which
exceeds 10 \% at $\lambda \sim 140$ nm and 50 \% at $\lambda < 120$ nm.

An array of empirical models that represent  
the different magnetic structures on the solar surface 
(e.g. sunspots, plage, network, etc.) needs to be considered
 to describe the variability of the solar spectrum throughout the solar cycle 
(see, e.g., Fligge, Solanki \& Unruh 2000, Fontenla et al. 1999), 
but at $\lambda > 200$ nm, a single
model is expected to be a reasonable approximation, given that
the vast majority of the  solar surface is typically free from  
regions with strong magnetic fields
(what is usually referred to as the 'quiet' Sun or the internetwork).

There is an extensive literature on the comparison of calculated and
observed solar UV fluxes. Most readers will remember 
the UV {\it missing opacity} problem, but the literature on this subject has
been sparse over the last decade. We first review recent
results, and then move on to describe our current efforts 
to improve models of the solar photosphere and 
compile updated opacities.

\section{Anybody said 'missing' opacity?}
\label{sec:2}

Early studies found too much UV flux in model atmosphere calculations
(Houtgast \& Namba 1968, Labs \& Neckel 1968, Matsushima 1968). 
Based on  a linelist from
semi-empirical calculations of atomic structure 
(Kurucz \& Peytremann 1975), completed with literature values, Kurucz  (1992) 
concluded that the problem was solved, but
his proposal was criticized by Bell et al. (1994), as high-resolution
observations did not confirm many of the predicted features. 

Bell, Balanchandran \& Bautista (2001) 
revisited this issue  armed with updated Fe I opacity from the
R-matrix calculations of Bautista (1997), concluding that the 
problem was significantly reduced, but still present. They found that 
if iron opacity was responsible
for the deficit, the new data could only account for half of the missing
opacity. More recently, we performed a similar study using Gaussian-averaged
photoionization cross-sections from the opacity project for elements with
atomic numbers 6--14 and scaled hydrogenic cross-sections 
for Fe I, arriving at the opposite
conclusion (Allende Prieto, Hubeny \& Lambert 2003a). 

Independently, Fontenla et al. (1999, see also Fontenla et al. 2006) 
used a combination of semi-empirical models to model the solar spectrum.  
They noticed an
opacity deficit around 410 nm. Nonetheless, the use of semi-empirical models 
whose temperature structure have been modified to reproduce 
observed fluxes, makes the discussion of absolute fluxes 
somewhat circular. Note also that the continuum metal opacities 
considered in these studies are outdated and neglect atomic iron.

There were several differences among the calculations of Balachandran
et al.  and ours. First of all, different model atmospheres were used:
a MARCS model versus an interpolated Kurucz (1993) solar model. 
A different solar surface composition
was adopted by the two groups. Most relevant, Balachandran et al. adopted 
$\log \epsilon$(Mg)$=7.44$ and $\log \epsilon$(Fe)$=7.55$, and we used
$\log \epsilon$(Mg)$=7.58$ and $\log \epsilon$(Fe)$=7.50$. Our higher
magnesium abundance can explain up to about 5 \% less flux in our calculations
at 400 nm and up to 20 \% shortwards of 300 nm (see Section \ref{sec:4}), 
but the difference between the iron abundances, although smaller,
 goes in the wrong direction.

Our calculations had (at least!) one prominent shortcoming: 
molecular opacity was neglected. We also made a mistake, 
including natural damping in L$\alpha$ 
too far from the transition's frequency. Mathematically, 
the natural damping contribution
to the Lorentzian wings of L$\alpha$ is
strong  enough to contribute very far, even into the optical.
Natural damping in Lyman alpha far from the transition 
frequency becomes in fact
Rayleigh scattering, and should be treated as such.

The opacity deficit, if any, has not been clearly linked 
to photoionization of
atomic iron, and the solar photospheric abundances of several major
elements have been systematically reduced over the last few
years (see  Asplund 2005, Asplund, Grevesse \& Sauval 2005). 
It is time to take a closer look at this issue.

\section{Revisiting the problem: opacities, equation of state, chemical
composition and model atmospheres}
\label{sec:3}

The problem of atmospheric structure, regardless of geometry, is
intrinsically coupled to the chemical composition of the star. The
relevant atomic and molecular opacities need to be accounted for, 
not only to predict accurately the spectrum shape, but also to
describe properly the energy balance, equation of state and, 
ultimately, the atmospheric structure (see the paper by Hubeny in this volume).
The chemical composition of the solar atmosphere, in turn, is determined
from spectral synthesis calculations based on a model atmosphere
computed for a given composition.
Thus, abundances, opacities, equation of state, and model atmospheres
are intrinsically coupled: changing one of these elements in isolation
may be meaningless.

Below, we briefly describe the main updates that we are implementing in
our calculations. 

\subsection{Abundances}

Over the last 7 years, a number of spectroscopic investigations 
of the solar chemical composition have significantly modified the standard values generally
adopted for the solar photosphere. The largest updates affect some of the
most abundant elements, such as carbon or oxygen (Allende Prieto,
Asplund \& Lambert 2001, 2002, Asplund et al. 2004, 2005), but minor 
changes affect also 
iron (Asplund et al. 2000b), silicon (Asplund 2000), or calcium 
(Asplund et al. 2005). The latter reference summarizes
these revisions, which are based on a new generation 
of three-dimensional time-dependent (non-magnetic) 
simulations of the solar surface.
Updates have also been made for heavier elements  
(Sneden \& Lawler 2005), albeit their
impact on the solar absolute fluxes is only marginal.

In our calculations, we have adopted the mixture proposed by Asplund et 
al. (2005). Note, however, that this compilation is not based on
a homogeneous analysis with a single model atmosphere and a uniform
protocol. The abundances for a number of elements are derived afresh, 
but for others 
it represents a critical evaluation of
new and old results, by different authors with various degrees 
of simplification, such as a strict adherence to LTE or the
adoption of NLTE corrections for some species, which are still unavailable 
or unreliable for many other,
in particular when it comes to 3D calculations. 

\subsection{Opacities}

After the widely-used photoionization cross-sections of Peach (1970), 
a significant improvement came with   
the calculations of atomic structure and opacities performed
by the international collaboration known as the Opacity Project
(Seaton et al. 1992). 
Until very recently, the Opacity Project (OP) provided 
two extreme products: cross-sections for each atomic state, or 
Rosseland mean opacities. For calculating synthetic fluxes, or model
atmospheres, one needs monochromatic opacities, but LTE codes
deal most comfortably with opacities per species, and do not need
detailed photoionization cross-sections for every single energy 
configuration. This situation has recently changed with the release 
of monochromatic opacities for each element as a function of temperature
and electron density (Seaton 2005), but the inconvenience of having to
deal with cross-sections has likely to do with the slow integration
of the OP data in astronomical codes.

We have implemented model atoms and ions for the most relevant
species for F-G-K-type atmospheres using the OP photoionization 
cross-sections (Allende Prieto et al. 2003b).
The data format follows the specifications for the NLTE 
model atmosphere code Tlusty (Hubeny \& Lanz 1995) and
the spectral synthesis code Synspec (Hubeny \& Lanz 2000).
As the computed energy levels are
relatively inaccurate, the location of the predicted resonances
(associated with two-electron autoionization; see the review article
by Sultana Nahar in this volume) in the cross-sections
is uncertain, and therefore we have smoothed them following the
prescription proposed by Bautista, Romano \& Pradhan (1998). 
These models continue to be updated periodically, and are 
publicly available\footnote{{\tt http://hebe.as.utexas.edu/at/}
and {\tt http://nova.astro.umd.edu/}}.

The OP calculations cover most elements from hydrogen through calcium,
but for iron ions have been superseded
by newer results from the Iron Project (see Bautista 1996, 1997, 
Bautista \& Pradhan 1997, Nahar \&  Pradhan 1996, 1999, and Nahar's paper 
in this volume). The 
distribution of data for Fe I and Fe II (the relevant iron ions for
late-type stellar atmospheres) through the Iron Project data base
is still patchy, but working in collaboration with Manuel Bautista and
Sultana Nahar, I have {\it translated} the data files to the same
format employed by the OP, and new model atoms for Tlusty/Synspec
have been produced.

\begin{figure}
\centering
\includegraphics[height=12cm,angle=90]{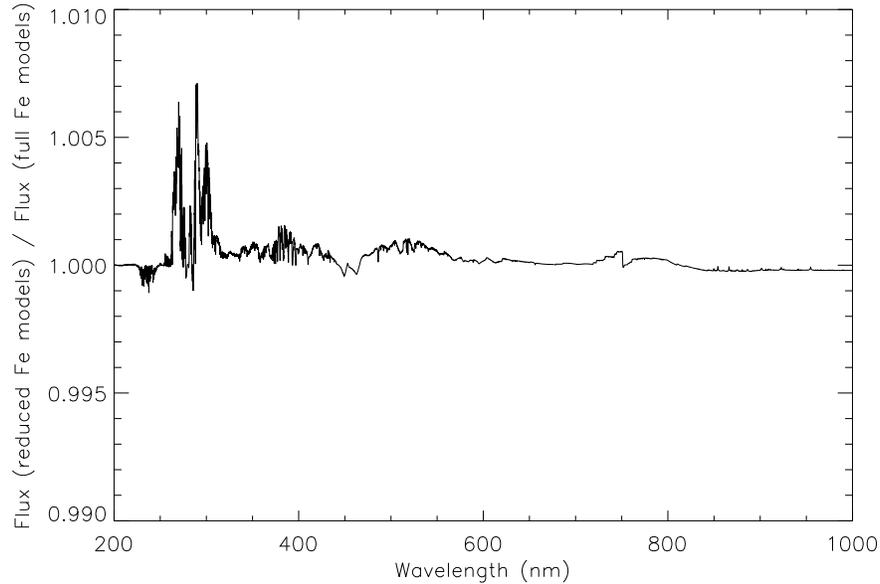}
\caption{Ratio of the solar irradiances computed with a simplified and
 full-blown  iron model atoms (Fe and Fe$^{+}$). The full-blown models account
 for the radiative opacity from more than  a thousand levels, while the
 boiled-down models include only about a hundred.}
\label{fig:1}       
\end{figure}

The Iron Project model ions are significantly larger
than those for lighter elements based on the OP data,
including of the order of 700 energy levels per ion. Assuming the 
relative populations
of levels with similar energies and the same quantum numbers L and S
are in equilibrium at a given temperature, it is possible to combine 
the cross-sections of these levels creating {\it super-levels}. The concept
of super-levels, 
introduced by Anderson (1989; see also Hubeny \& Lanz 1995), 
can be exploited to effectively
reduce the complexity of the 
opacity calculations, as well as to speed up the solution
of the rate equations in NLTE problems. For a solar-like atmosphere,
using this simplification for Fe I (assuming $T=5000$ K) 
and for Fe II ($T=7000$ K), leads to errors in the computed absolute flux 
less than 1 \% when the size of the model atoms is reduced tenfold,
as shown in Fig. 1.

\subsection{Equation of state}

By adopting a model atmosphere that has been precalculated,
all relevant thermodynamical quantities are readily available
as a function of the location in the atmosphere. As we discussed
above, the input atomic and molecular data, as well as the
abundances, will determine the resulting structure and
energy flux, but some quantities, such as the emergent flux, are
expected to be more sensitive to small variations in some of the
basic inputs than others, such as the thermal structure 
of the model atmosphere.

We have explored the effect of small changes in the input
chemical composition on the emergent fluxes by considering
the  thermal atmospheric structure fixed (see Section \ref{sec:4}). 
Under this
approximation, we still recompute consistently the electron
density and solve the molecular equilibrium. This step involves
a major upgrade from our earlier calculations in order to 
consider the presence of molecules, their impact on the electron 
density, and ultimately on the atomic species 
(I. Hubeny, private communication). 
To this purpose, the most recent versions of Synspec include
routines kindly provided from U. J$\o$rgensen. Both
atomic and molecular partition functions are adopted from Irwin (1981
and private communication), while other molecular data are from
Tsuji (1973).

\subsection{Model atmospheres}

As argued above, computing absolute fluxes involves solving
consistently the problem of atmospheric structure and calculating 
the radiation field for any given set of abundances. 
We are using a NLTE model atmosphere code, 
but including in detail all the relevant
sources of opacity for late-type atmospheres and  
accounting for departures from LTE simultaneously is a massive problem. 
On the other hand, mild or no departures from LTE are expected for many
atomic and molecular species. Thus, we are working towards a hybrid
scheme where the contribution to the opacity for most species is
computed in LTE and stored in a look-up table, while only the
most relevant ions are considered in NLTE. 

We have already mentioned recent updates in the solar
photospheric abundances associated with a new kind of model atmospheres 
based on 3D hydrodynamics.
Surface inhomogeneities, in particular solar granulation, 
may have an important effect on the absolute 
flux emerging from the solar surface.  
Radiative transfer solvers for 3D 
are typically ready to handle only simple line opacities: 
one line profile or a few. 
Computing absolute fluxes, especially in the UV domain,
requires including very large number of overlapping atomic and
molecular transitions, in addition to detailed metal photoionization
cross-sections. To this goal, a new radiative transfer code has
been developed by L. Koesterke  (private communication), 
able to consider full-blown opacities, including electron and 
Rayleigh scattering. 

Koesterke et al. (2007) find that the solar 3D model by Asplund et al. (2000a) 
performs similarly to 1D models regarding limb darkening in the continuum, 
despite  a simplified description of the radiation field. 
In addition, the same model vastly outperforms 1D models 
regarding line formation, and in particular the 
center-to-limb variation of line profiles. 
The ability of 
3D models to match the solar limb darkening had been put
into question by Ayres et al. (2006). Based on tests using a horizontal-
and time-averaged structure from the simulations by Asplund et al. (2000a),
these authors predicted a dramatic failure of the new models.
The more rigorous calculations by Koesterke et al. 
show that the limb-darkening of
a three-dimensional model is very different from that of 
a 1D model derived by taking the average over surfaces 
with constant vertical optical depth. 
The effects 
of surface convection on the absolute solar fluxes are currently being
investigated with the new radiative transfer code.

\section{The role of chemical composition: a seven-pipe problem}
\label{sec:4}

Comparing absolute solar fluxes predicted by
model atmospheres with observations usually involves adopting 
a standard set of chemical abundances, but can we consider the chemical 
composition as a fixed set of parameters? The recent 
revisions for carbon and oxygen, together with the typical
error bars still quoted in solar abundance studies, which sometimes 
exceed 0.1 dex, suggest that the answer is NO.

\begin{figure}
\centering
\includegraphics[height=12cm,angle=90]{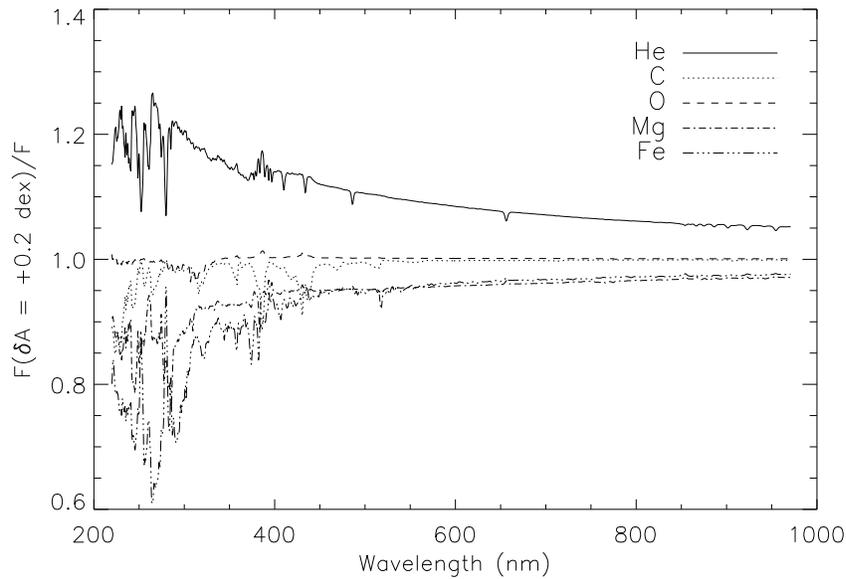}
\caption{Relative variations in the solar surface flux emergent from a 1D solar
model atmosphere resulting from changes in the adopted chemical composition.
The atmospheric structure (the run of temperature versus mass column density) 
is considered constant in these calculation.}
\label{fig:1}       
\end{figure}

Only a few elements can make an important impact on the computed
solar fluxes: directly through 
contributed opacity, or indirectly, by their effect on the atmospheric
structure or  the number of free electrons they release through
ionization. We have calculated, using a solar Kurucz model, the 
effect of changing the abundances of the most relevant elements 
on the solar spectrum. The results of 0.2 dex variations in the
X/H ratios, where X is  He, C, O, Mg, and Fe, are shown in Fig. 2.
Ca, Si, and some iron peak elements can also have an effect.

H$^{-}$ dominates the continuum opacity in the solar
optical and infrared. In the blue and UV, atomic iron and magnesium
contribute significant continuum opacity through photoionization, and
iron also provides abundant line opacity.  At wavelengths shorter
than 200 nm, aluminum and silicon need to be considered as well (but 
see comments in \S \ref{sec:1}).
Molecules, mainly CH, CO, and OH dominate relatively narrow
bands of the optical, IR and UV solar spectrum. Besides H, at least iron, 
magnesium and silicon, are significant contributors to the
pool of free electrons, which has a tremendous impact on the continuum
opacity as the number density  
of free electrons is smaller than that of hydrogen atoms 
and therefore controls the formation of H$^{-}$.

At first sight, the impact of changing the helium abundance in Fig. 2
may be a surprise. This is truly an indirect effect: as all abundances are
normalized to H and He is very abundant, N(He)/N(H) $\sim 0.07$, an increase
in He/H involves a significant reduction in N(H), and consequently
in the atomic hydrogen opacity, which results in an increased irradiance.
Fortunately, the solar He/H ratio is known precisely from helioseismology.

Inspection of Fig.2, considering that the observed solar absolute fluxes  
are likely accurate to a level of $\sim 1$ \% or better, 
indicates that the current
uncertainties in the  chemical composition of the solar surface may
be a dominant source of error in the flux calculations. 
This situation is similar to the case of the 
predicted solar neutrino fluxes! (Bahcall \& Serenelli 2005).

\section{Conclusions}
\label{sec:5}

Observations of the solar angular diameter by different authors and methods
still show rather significant discrepancies (see, e.g., Basu 1998, 
and the references
discussed by Wittmann \&  Neckel 1996), and probably a poorly-understood
time variation. Nonetheless, this quantity is
known with a relative accuracy many orders of magnitude higher than for
any other star, opening the possibility to compare detailed observed
absolute fluxes with the predictions from model atmospheres to learn
about physics and astronomy. 

So far, no assessment has been made of the potential impact of the
new generation of the 3D hydrodynamical model atmospheres on the
computed solar irradiance, but the presence of inhomogeneities
introduced by convective overshooting could alter the solar
spectrum significantly. Computing fluxes
from 3D models involves 
3D radiative transfer; using horizontally-averaged structures to explore
3D models  
is bound to lead to erroneous conclusions. The availability of 
several detailed hydrodynamical simulations of the solar
surface (e.g., Asplund et al. 2000a, Wedemeyer et al. 2004, V\"ogler
et al. 2005) contrasts with the scarcity of detailed radiative transfer
using them.

Computing absolute fluxes is more demanding than relative values,
and much more sensitive to input values such as the adopted chemical
composition. 
Modern opacities should be employed, in particular computed
state-of-the-art photoionization cross-sections for atomic
iron, magnesium, aluminum, and silicon, as well as 
line opacity from the most important diatomic molecules.
Our tests indicate the need for fully consistent
calculations in order to disentangle the impact of changes
in composition  and input micro-physics. The blue and UV fluxes of the Sun
are particularly sensitive to the abundances of hydrogen, carbon, oxygen,
magnesium, aluminum, silicon, calcium and iron.
Our preliminary results  hint that the uncertainties in the composition of  
the solar atmosphere may be a dominant source of error in    
predicting the radiation output of the Sun.

{\it Acknowledgments.} It is my pleasure to recognize 
significant contributions to
this work 
from  Martin Asplund, Manuel Bautista, Lars Koesterke, 
Sultana Nahar,  David Lambert, Thierry Lanz, 
and in particular Ivan Hubeny. I thank Emanuele Bertone, Miguel Ch\'avez,
Lino Rodr\'{\i}guez-Merino y Daniel Rosa-Gonz\'alez for their kind 
hospitality. Support from NASA (NAG5-13057, NAG5-13147) is 
thankfully acknowledged.



\printindex
\end{document}